# Enhanced spin ordering temperature in ultrathin FeTe films grown on a topological insulator

Udai Raj Singh[1], Jonas Warmuth[1], Anand Kamlapure[1], Lasse Cornils[1], Martin Bremholm[2], Philip Hofmann[3], Jens Wiebe[1] and Roland Wiesendanger[1]

[1]*Department of Physics, University of Hamburg, D-20355 Hamburg, Germany*

[2]*Department of Chemistry, Interdisciplinary Nanoscience Center (iNANO), Aarhus University, DK-8000 Aarhus C, Denmark*

[3]*Department of Physics and Astronomy, Interdisciplinary Nanoscience Center (iNANO), Aarhus University, DK-8000 Aarhus C, Denmark*

E-mail: usingh@physnet.uni-hamburg.de and jwiebe@physnet.uni-hamburg.de

# Abstract

We studied the temperature dependence of the diagonal double-stripe spin order in one and two unit cell thick layers of FeTe grown on the topological insulator $Bi_2Te_3$ via spin-polarized scanning tunneling microscopy. The spin order persists up to temperatures which are higher than the transition temperature reported for bulk $Fe_{1+y}Te$ with lowest possible excess Fe content $y$. The enhanced spin order stability is assigned to a strongly decreased $y$ with respect to the lowest values achievable in bulk crystal growth, and effects due to the interface between the FeTe and the topological insulator. The result is relevant for understanding the recent observation of a coexistence of superconducting correlations and spin order in this system.

Superconductivity in bulk iron pnictides and chalcogenides, similar to cuprates, emerges upon the suppression of magnetic order in the parent compounds via doping, signifying an intimate relation between magnetism and superconductivity [1, 2, 3]. In the parent compounds, iron pnictides are known to have a collinear antiferromagnetic (AFM) order with an ordering wave vector ($\pi$, 0) [4], while iron chalcogenides exhibit a diagonal double-stripe (DDS) AFM order with a wave vector ($\pi$, $\pi$), i.e. rotated by 45° with respect to the Fermi-nesting wave vector between hole- and electron-pockets [5, 6, 7]. In most cases of doped chalcogenides in the superconducting state in which the magnetic order is weakened, inelastic neutron scattering experiments, however, show the presence of a spin resonance mode at the wave vector ($\pi$, 0), signifying a $s_\pm$ type pairing or a sign change in the superconducting gap [4, 5, 8, 9]. Hence, the role of the spin degree of freedom becomes decisive for understanding the mechanism of high-$T_c$ superconductivity in these materials. A real-space visualization of spin-dependent phenomena on the atomic scale in dependence of the doping and temperature can thus provide a new dimension towards an understanding of high-$T_c$ superconductivity. A suitable technique for this purpose is spin-polarized scanning tunneling microscopy (SP-STM) which has widely been utilized in studying magnetic nanostructures and spin-dependent phenomena of metallic materials on the atomic scale [10].

While superconductivity in chalcogenide materials such as $Fe_{1+y}Te_xSe_{1-x}$ has been extensively investigated by spin-averaging STM in the past few years [11, 12, 13], there are only few results using SP-STM for these materials [14, 15, 16, 17, 18]. The DDS spin order in the parent compound $Fe_{1+y}Te$, emerging around the Néel temperature ($T_N \sim$ 65 K) from the paramagnetic state [6, 7, 19, 20, 21], has been recently confirmed by SP-STM at low temperatures ($T$) [14, 15, 16, 18]. In the regime of low excess Fe content $y$, $T_N$ generally strongly increases with decreasing $y$ [19], with reported maximum values of $T_N$ of 72 K [20, 21], which is limited by the lowest achievable values of $y$ in conventional bulk crystal growth.

A DDS spin order has also been found in the cases of one and two unit cell (1-2 UC) thick FeTe films grown on the topological insulator (TI) $Bi_2Te_3$ [17, 18]. Most notably, for these thin films of FeTe, superconductivity coexists with the spin order at $T$ < 6 K [17, 22, 23]. The system is particularly interesting, as interfaces of s-wave superconductors with TIs are predicted to resemble spinless $p_x$-$p_y$ superconductors that may support Majorana bound states in vortices [24]. Although changes of $T_N$ of the thin film system with respect to the bulk material are expected because of a possible decrease in $y$ as compared to the lowest achievable values in cystal growth [19], strain effects [25], or charge transfer between the FeTe and the TI [26], the magnetic phase transition from the DDS spin ordered into the paramagnetic phase has not been studied so far on the local scale. Here, we present a temperature-dependent SP-STM investigation of 1-2 UC thick FeTe/$Bi_2Te_3$ films in the temperature range from 1 K to 80 K. We observe an enhancement in $T_N$ as compared to the maximum values reported for bulk materials and discuss possible reasons for this phenomenon.

For the growth of 1-2UC thick layers of FeTe on $Bi_2Te_3$(0001), 3.5 monolayers of Fe were deposited with an e-beam evaporator onto the *in-situ* cleaved surface of $Bi_2Te_3$ at room temperature, followed by annealing at 300°C for 2 h [17, 18]. SP-STM measurements were carried out in a variable-temperature STM in the temperature range from 25 K to 80 K [27, 28, 29] and in a SPECS STM at 1.1 K [30]. We used bulk Cr tips which were cleaned *in-situ* by field emission and voltage pulses against a W(110) single crystal prior to the SP-STM measurements [17]. (SP-) STM images were recorded in constant current mode with a tunneling current of $I_t$ at a bias of $V_b$ applied to the sample.

Figure 1(a) illustrates an overview STM image of the sample showing the growth of FeTe islands of different lateral sizes and heights on the $Bi_2Te_3$ substrate. A detailed atomic scale investigation of such types of FeTe films, prepared using the same growth process, was carried out at low $T$ [17]. The line profile in Fig. 1(b) across four selected islands reveals four distinct heights of 6.4 Å, 12.8 Å, 8.0 Å, and 14.4 Å above the $Bi_2Te_3$ surface. Comparing these heights to the height of the FeTe UC, $c \sim 6.3$ Å [7], we can assign them to one and two UC thick FeTe islands on top of the substrate, and to one and two UC thick FeTe islands on an FeTe layer embedded in the topmost quintuple layer (QL) of the $Bi_2Te_3$, respectively [see Fig. 1(b)]. Note, that we cannot unambiguously determine the thickness of the embedded FeTe layer from our STM images. While the $Bi_2Te_3$ substrate is probably terminated by a complete QL below the one UC and two UC islands, it is terminated by an incomplete QL for the case of the embedded FeTe islands. We therefore disregard the embedded layers in the following and focus on the FeTe layers which are grown on the intact TI. Figure 1(c) shows an atomically resolved STM topography from a substrate region revealing the expected hexagonal symmetry, which is further supported by a hexagonal pattern of Bragg peaks appearing in the fast Fourier transformation (FFT) as provided in Fig. 1(e) (Fourier amplitudes). On the other hand, the atomically resolved STM images obtained on top of the islands show the presence of the expected square lattice of surface Te atoms as presented in Fig. 1(d). Correspondingly, the FFT in Fig. 1(f) displays Bragg peaks associated with the lattice vectors ($\mathbf{q}_{Te}^a$ and $\mathbf{q}_{Te}^b$) appearing as a square pattern. The measured lattice constant is $a_0 \approx 3.78$ Å equivalent to the lattice constant of FeTe [7]. Note, that the surface of the FeTe film is free from the presence of excess Fe atoms which are clearly visible as adatoms after the cleavage of bulk crystals [14]. Thus, our FeTe films provide an opportunity to study a pure stoichiometric composition of 1-2 UC thick layers of FeTe, being unavailable in the case of bulk materials [7].

Figure 2(a) and 2(b) illustrate atomically resolved topographic STM images taken at $T = 28.5$ K and 66.7 K. Note, that there are no signs of any surface reconstruction or periodicities other than that of the surface Te lattice of $a_0$ in the whole investigated temperature range. Consistently, the FFTs of the STM topographies in Figs. 2(c) and 2(d) show only the Bragg peaks and no additional features associated with the DDS spin order of the FeTe thin films [17, 18]. This indicates that, even though the used bulk tip material is magnetic, the particular tip apex used for these images has no spin-polarization. The absence of any $2a_0$ modulation in such spin-averaging STM images is in contrast to an earlier STM study on FeTe(001) films claiming the existence of a periodic charge order induced modulation of $2a_0$ at $T < 60$ K [31].

In order to obtain magnetic contrast in the STM images, we deliberately altered the tip apex by scanning at bias voltages up to $V_b = 2$ V with tunnelling currents of $I_t = 5$ nA in order to prepare a spin-polarized tip. Changes of the tip apex were further provoked by scanning across the step edge between FeTe islands and the substrate, as well as via field emission. Using such spin-polarized tips, Fig. 3(a) shows constant-current SP-STM data taken on a one UC thick layer of FeTe at $T = 1.1$ K. The magnetic contrast corresponding to the DDS spin order [6, 7] with a periodicity of $2a_0$ is clearly visible, which is further validated by the presence of two additional peaks at $\mathbf{q}_{AFM} = \pm\frac{1}{2}\mathbf{q}^a_{Te}$ in the FFT marked by a red circle in Fig. 3(e), in agreement with the results reported earlier for the cases of ultrathin FeTe films [17, 18] and bulk systems [14, 15, 16, 17]. Upon increasing $T$, the magnetic contrast due to the DDS order is visible at 48.4 K [Fig. 3(b)], 67.1 K [Fig. 3(c)], and 73.2 K [Fig. 3(d)]. In their FFTs [Figs. 3(f)-(h)], the peaks associated with the DDS order are seen all the way up to 73.2 K. Note, that the appearance of the $2a_0$ periodic spin-contrast in the SP-STM images non-systematically changes between two different types: for the first type [Figs. 3(a,c,d)], the maximum of the contrast is located in between the rows of the Te atoms, i.e. every second trench between the Te rows in $\mathbf{q}^b_{Te}$ direction appears deeper, while all Te atoms have the same apparent height. For the second type [Fig. 3(b)], the maximum of the contrast is located on top of the Te atoms, i.e. every second Te row in $\mathbf{q}^b_{Te}$ direction appears dimmer, while all trenches have the same apparent depth. While, usually, either of these two pure types of spin contrasts were observed, some tips show a mixture, as visible, e.g., in the SP-STM image in [Fig. 4(b)]. We have also observed that the spin-contrast can change from one to the other type by a change of the tip apex while taking an SP-STM image (not shown). Similar contrast changes have been observed in Ref. [16], and were studied as a function of bias voltage. We assign them to an energy- and tip-dependent strength and orientation of the spin-polarization, which may additionally depend on $T$. Since the tip apex usually changed between the measurements shown in Fig. 3 (as well as in Fig. 4 below), we cannot determine whether the observed changes in the strength of the spin contrast and in the according intensities of the $\mathbf{q}_{AFM}$ peaks in the FFTs in Fig. 3 are due to the changes in the tip's spin-polarization described above, or due to temperature induced fluctuations of the DDS spin-order. However, since the periodicity of the spin-contrast does not change, we can conclude, that a long range DDS spin order persists at least up to 73.2 K in the UC thin film on $Bi_2Te_3$. For higher temperatures, we did not succeed in observing the $2a_0$ contrast, supposedly because the film turns into its paramagnetic state.

In the case of a two UC thin FeTe film, a similar magnetic contrast related to the DDS spin order has been clearly resolved at 1.1 K [Fig. 4(a)], 30.3 K [Fig. 4(b)], and 70.2 K [Fig. 4(c)]. FFT maps correspondingly show the peaks related to the DDS order [Figs. 4(e-g)]. Most notably, the $2a_0$ periodic contrast is still faintly observable at 79.3 K [Figs. 4(d,h)], indicating that the DDS spin order in the two UC thin FeTe film on $Bi_2Te_3$ persists up to temperatures larger than the highest bulk $T_N$ ~ 72 K [20, 21].

In the following, we discuss possible reasons for the increased $T_N$ in our thin FeTe films grown on the TI $Bi_2Te_3$ as compared to the bulk system. In principle, there are four obvious effects. (i) For the bulk system, a decrease in $y$ leads to a considerable increase in $T_N$ [19]. Using a linear extrapolation of the $T_N(y)$ data of Ref. [19] to $y = 0$, $T_N$ ~ 85 K would be achievable in the bulk system for $y = 0$. Considering a zero $y$ of our thin films as suggested by the lack of Fe adatoms, the increased $T_N$ is naturally explained by this effect. (ii) The interaction of the FeTe thin film with the substrate leads to strain effects, which are also known to have a strong effect on $T_N$ [25]. (iii) ARPES measurements of the band structure of similar samples as the ones investigated here showed a considerable transfer of electrons from the FeTe thin film to the TI [26], which could as well affect $T_N$. (iv) The electronic structure of the thin film can be slightly different form the bulk due to quantization effects in the perpendicular direction, which can change $T_N$ [26]. We note that heterostructures of various materials show properties of one and two-monolayers which are significantly different from the corresponding bulk systems. Examples are the superconducting transition temperature in a single layer of FeSe on $SrTiO_3$(001) [32], the absence of a charge density wave in a single layer 2H-$TaS_2$ [33] the enhanced charge density wave transition temperatures in a single layer of $TiSe_2$ on 6H-SiC(0001) [34] and in a monolayer of $NbSe_2$ [35], and the room temperature ferromagnetism in one and two QLs of $Bi_2Se_3$ on EuS [36]. For the case of $Bi_2Se_3$/EuS, the surface states of the TI, which are spin-momentum locked due to strong-orbit interaction, play an essential role in enhancing the magnetic ordering temperature at the interface of this system [36]. Whether or not similar effects play an important role for the enhancement of $T_\text{N}$ observed here is an interesting question remaining for future investigations.

In conclusion, we have demonstrated by SP-STM that the DDS spin orders in one and two UC thick layers of FeTe on $Bi_2Te_3$ persist up to 73 K and 79 K, respectively. Consequently, the spin order transition temperatures in our FeTe thin films are higher than the maximum $T_\text{N}$ found for $Fe_{1+y}$Te bulk materials with lowest possible $y$. We attribute this favoring of the spin order to the strongly decreased excess Fe content as compared to the bulk system and to interface effects between the FeTe film and the TI substrate.

# Acknowledgments

We thank Elena Vedmedenko, Julian Hagemeister, Bin Shao, and Tim Wehling for fruitful discussions. We acknowledge funding by the ERC via the Advanced Grant ASTONISH (No. 338802). J. Wi., L. C., and P. H. acknowledge support by the DFG via the priority program SPP1666 (grant nos. HO 5150/1-2 and WI 3097/2). This work was supported by VILLUM FONDEN via the Centre of Excellence for Dirac Materials (Grant No. 11744). M. B. acknowledges support from the Center for Materials Crystallography funded by the Danish National Research Foundation (DNRF93).

# Figures

**FIG.1.** (a) Overview constant current STM image of the sample taken at $T = 28$ K showing FeTe islands of four different heights above the $Bi_2Te_3$(0001) surface ($V_b = 100$ mV, $I_t = 100$ pA). The four different FeTe island types and the substrate area are indicated. (b) Height profile taken along the line in (a) passing one of each of the four FeTe island types with distinct height as indicated. (c,d) Atomically-resolved constant-current STM images (c) from a substrate region and (d) on top of an island (c: $T = 1.1$ K, $V_b = 100$ mV, $I_t = 200$ pA; d: $T = 1.1$ K, $V_b = 200$ mV, $I_t = 100$ pA). The measured lattice constant of FeTe ($a_0 \approx 3.78$ Å) is indicated in (d). (e,f) FFTs of (c,d), respectively. The primary FeTe Bragg peaks are indicated by blue and yellow circles.

**FIG.2.** (a,b) Atomically resolved constant current STM images of one UC thick FeTe on $Bi_2Te_3$(0001) taken at the indicated temperatures (a: $V_b = 20$ mV and $I_t = 2$ nA; b: $V_b = -30$ mV, $I_t = 4$ nA). (c,d) FFTs of (a) and (b), respectively, with the primary Bragg peaks corresponding to the square lattice of Te atoms ($\mathbf{q}_{Te}^a$ and $\mathbf{q}_{Te}^b$) indicated by blue and yellow circles.

**FIG.3.** (a-d) Constant-current SP-STM images of one UC thick FeTe on $Bi_2Te_3$(0001) taken at $T$ as indicated (a: $V_b = 100$ mV, $I_t = 100$ pA; b: $V_b = 10$ mV, $I_t = 3$ nA; c: $V_b = 20$ mV, $I_t = 2$ nA; d: $V_b = 60$ mV, $I_t = 2$ nA). The double periodicity of the spin-contrast ($2a_0$) is indicated. The inset in (a) shows the top view of a ball-model of the FeTe lattice with the DDS spin order of the Fe atoms indicated by arrows [Te (u) = upper atom and Te (d) = lower atom]. (e-h) FFTs of (a-d). The FFT peaks corresponding to the square lattice of Te (u) atoms ($q^a_{Te}$ and $q^b_{Te}$) and to the DDS order ($q_{AFM}$) are indicated by blue, yellow, and red circles, respectively.

**FIG.4.** Constant-current SP-STM images of two UC thick FeTe on $Bi_2Te_3$(0001) acquired at $T$ as indicated (a: $V_b$ = 100 mV, $I_t$ = 100 pA; b: $V_b$ = -10 mV, $I_t$ = 5 nA; c: $V_b$ = 80 mV, $I_t$ = 5 nA; d: $V_b$ = 40 mV, $I_t$ = 3 nA). (e-h) Corresponding FFTs of (a-d) with indicated FFT peaks (same as in caption of Fig. 3).

# Figure 1

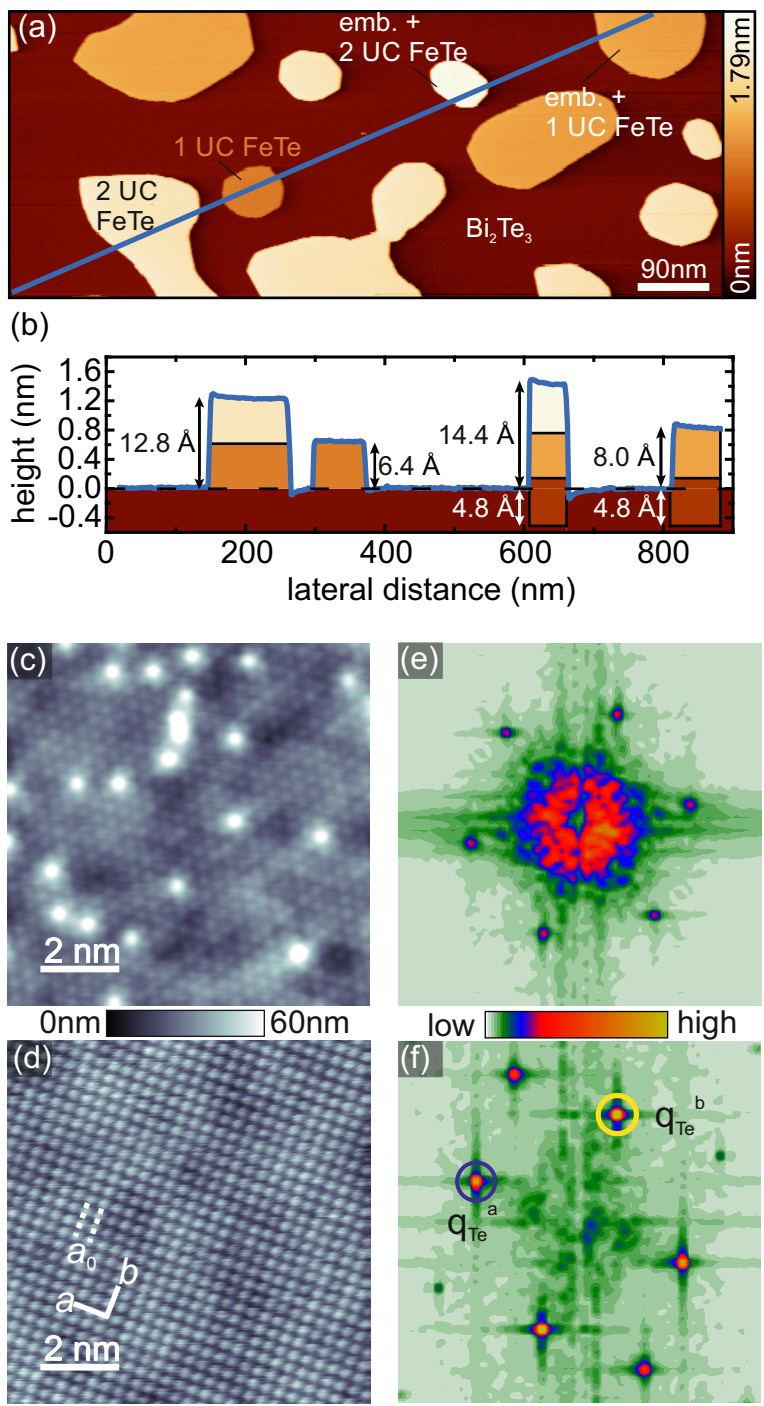

# Figure 2

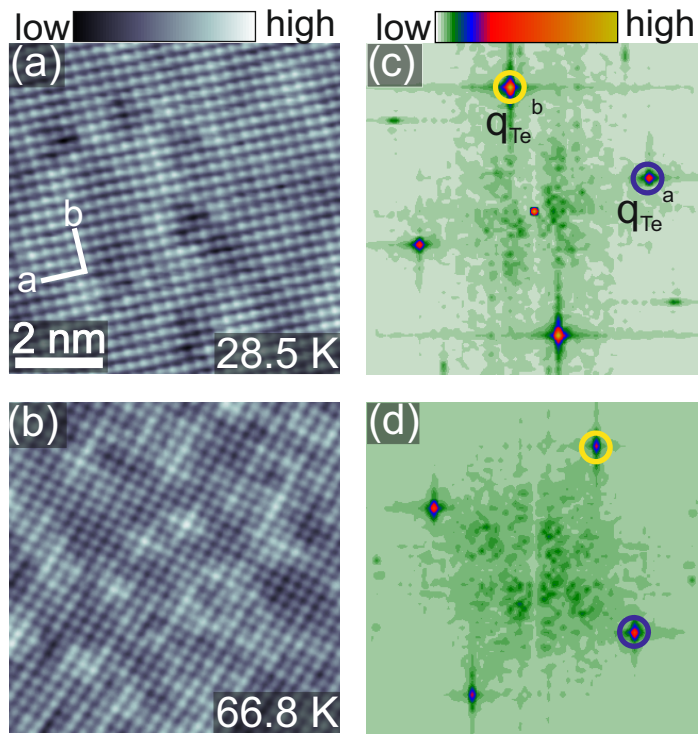

# Figure 3

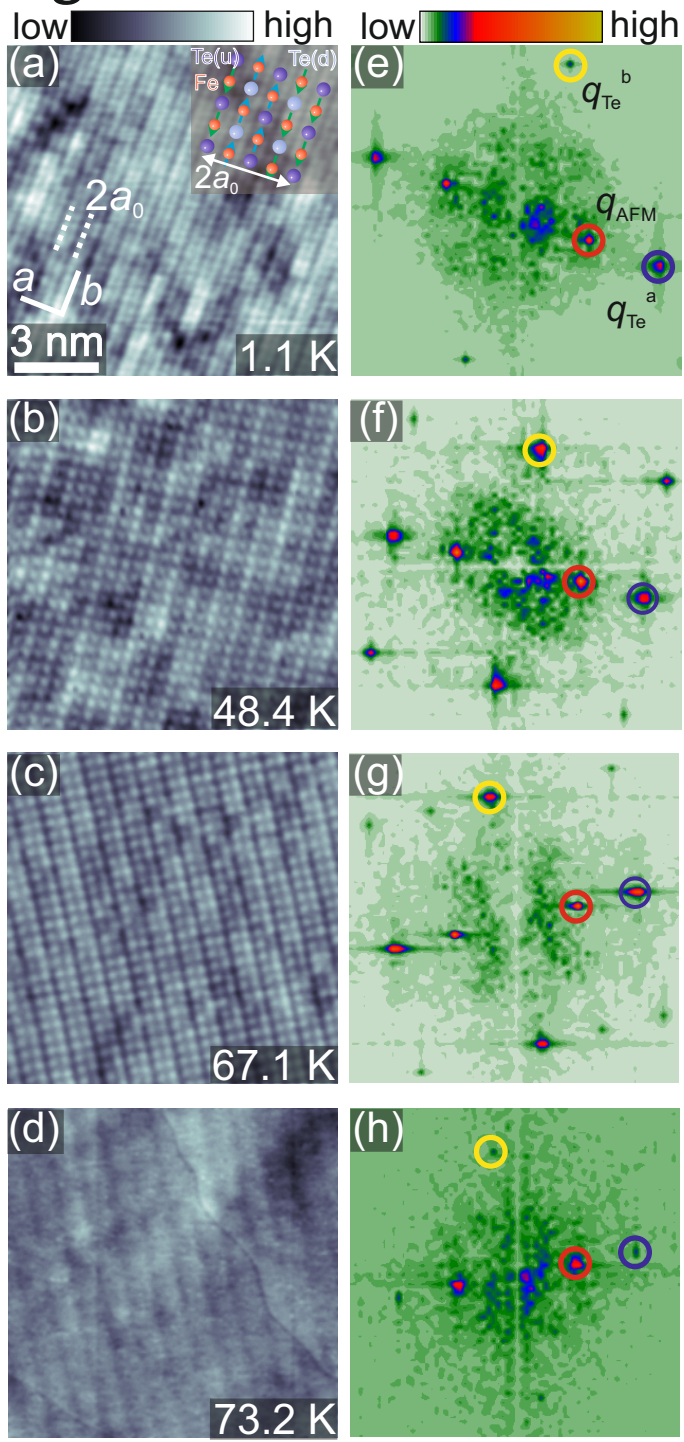

# Figure 4

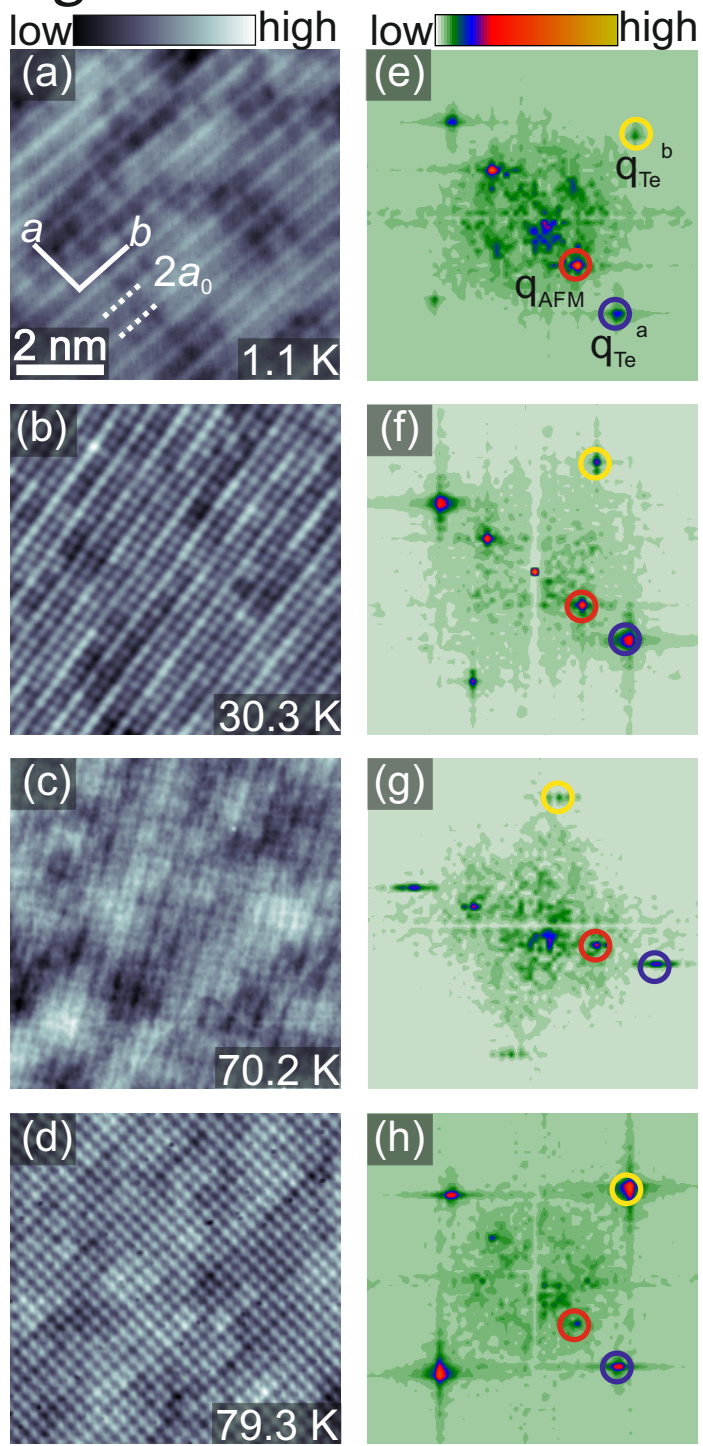